\documentstyle[12pt,epsf]{article}

\def \dfrac #1#2 {\displaystyle\frac{#1}{#2}}
\def\be{\begin{eqnarray}}
\def\ee{\end{eqnarray}}
\def\bq{\begin{equation}}
\def\eq{\end{equation}}
\def\ben{\begin{enumerate}}\def\een{\end{enumerate}}

\def\pr{Phys. Rev. }
\def\np{Nucl. Phys. }\def\pl{Phys. Lett. }

\def\roughly#1{\mathrel{\raise.3ex\hbox{$#1$\kern-.75em%
\lower1ex\hbox{$\sim$}}}}

\begin{document}
\begin{titlepage}

\hfill {FTUV 98-8; IFIC 98-8}


\vspace{.2cm}

\begin{center}
\ \\
{\Large \bf Polarized structure functions in a constituent quark scenario $\dagger$}
\ \\
\vspace{.5cm}
{Sergio Scopetta and Vicente Vento$^{(a)}$}

{\it Departament de Fisica Te\`orica, Universitat de Val\`encia}

{\it 46100 Burjassot (Val\`encia), Spain}
 
            and

{\it (a) Institut de F\'{\i}sica Corpuscular, Consejo Superior de 
Investigaciones Cient\'{\i}ficas}
\vskip 0.2cm            
                 and
\vskip 0.2cm                
{Marco Traini}

{\it Dipartimento di Fisica, Universit\`a di Trento, I-38050 Povo (Trento),
Italy} 

{\it and Istituto Nazionale di Fisica Nucleare, Gruppo Collegato di Trento}

\end{center}
\vskip 0.5cm
\centerline{\bf Abstract}
\vskip 0.2cm
Using a simple picture of the constituent quark as a
composite system of point-like partons, we construct the 
polarized parton distributions by
a convolution between constituent quark momentum distributions
and constituent quark structure functions.
Using unpolarized data to fix the parameters we achieve good 
agreement with the polarization experiments for the proton, while not so
for the neutron. By relaxing our 
assumptions for  the sea distributions,  
we define new quark functions for 
the polarized case, which 
reproduce well the proton data and are in better agreement with the 
neutron data.    

When our results are compared with similar calculations using non-composite 
constituent quarks the accord with the experiments of the present scheme 
is impressive. We conclude that, also in the polarized case,
DIS data are consistent with a low energy scenario
dominated by composite constituents
of the nucleon.

\vskip 0.5cm
\leftline{Pacs: 12.39-x, 13.60.Hb, 14.65-q, 14.70Dj}
\leftline{Keywords:  hadrons, quarks, gluons, evolution, parton distributions, 
structure functions.} 
\vspace{.5cm}

{\tt
\leftline {scopetta@titan.ific.uv.es}
\leftline{vicente.vento@uv.es}
\leftline{traini@science.unitn.it}
}
\vspace{0.5cm}
\noindent{\small$\dagger$Supported in part by DGICYT-PB94-0080 
and TMR programme of the European Commission ERB FMRX-CT96-008}.
\end{titlepage}

\section{Introduction}
At low energies, the idea that baryons are made up of three constituent
quarks and mesons of a  (constituent) quark-antiquark pair \cite{Gell-Mann},
the so called naive quark model, accounts for a large number of experimental
observations \cite{Quark-Model}. 
Soon after the formulation of the naive quark model, deep inelastic scattering
of leptons off protons was explained in terms of pointlike constituents named
partons \cite{Feynman}. The birth of $QCD$ \cite{Fritsch} and the proof that it is
asymptotically free \cite{Gross-Wilczek-Politzer} set the framework for an
understanding of the deep inelastic scattering phenomena beyond the Parton
Model \cite{Muta-Field-Yndurain}. However, the perturbative approach to $QCD$
does not provide absolute values for the observables. The description based
on the Operator Product Expansion ($OPE$) and the $QCD$ evolution requires the
input of non-perturbative matrix elements. We have developed an approach which 
uses  model calculations for the non-perturbative 
matrix elements \cite{Traini}. 
Moreover, in order to relate the constituent quark with the current partons of
the theory a procedure, hereafter called ACMP, has been applied
\cite{Altarelli1,Scopetta}.

In our approach \cite{Scopetta} constituent quarks are effective particles 
made up of point-like partons (current quarks (antiquarks) and gluons), interacting by a
residual interaction described as in a quark model. The hadron structure
functions are obtained by a convolution of the constituent quark model
wave function with the constituent quark structure function. 

The procedure has been recently used to estimate the pion structure function
\cite{Altarelli2} and the unpolarized proton one with 
success \cite{Scopetta} . We have shown that $DIS$ data are consistent with a 
low energy scenario dominated by composite, mainly non-relativistic 
constituents of the nucleon. In here we extend our analysis to the polarized 
$g_1$ structure function.

Summarizing: In section 2 we will review briefly the formalism for the 
unpolarized case to set the ground for the discussion and present the 
generalization for the  polarized one. Section 3 will contain the comparison 
with  the  data of the calculated structure functions.  Finally, Section 4 
will contain the conclusions and outlook.

\section{The theoretical framework}

In our picture the constituent quarks are
themselves complex objects whose structure functions are described by a set of
functions $\Phi_{ab}$ that specify the number of point-like partons of type $b$,
which are present in the constituents of type $a$ with fraction $x$ of its total
momentum \cite{Altarelli1,Scopetta}. In general $a$ and $b$ specify all the relevant quantum numbers of the
partons, i.e., flavor and spin. The unpolarized case for the proton was discussed
in detail in ref.\cite{Scopetta}. We proceed to a short review of the
description in order to  set the ground for the study of the polarized 
structure function $g_1$.

The functions describing the nucleon parton distributions omitting spin degrees
of freedom are expressed in terms of the independent $\Phi_{ab}(x)$ and of the
constituent probability distributions $u_0$ and $d_0$, at the hadronic scale
$\mu_0^2$ \cite{Traini}, as
\bq
f(x,\mu_0^2) = \int_x^1 \frac{dz}{z}[u_0(z,\mu_0^2)\Phi_{uf}(\frac{x}{z},\mu_0^2) +
d_0(z,\mu_0^2)\Phi_{df}(\frac{x}{z},\mu_0^2)]
\eq
where $f$ labels the various partons, i.e., valence quarks ($u_v,d_v$), sea
quarks ($u_s,d_s,s$), sea antiquarks ($\bar{u},\bar{d},\bar{s}$) and gluons $g$.

The different types and functional forms of the structure functions for the
constituent quarks are derived from three very natural assumptions, extensively
discussed and theoretically motivated \cite{Altarelli1}. Although these ideas
were proposed before $QCD$ was fully developed, they can be easily transported
to it and result in:

\begin{itemize} 
\item[  i)]The point-like partons are determined by $QCD$, therefore, quarks, 
antiquarks and gluons;
\item[ ii)] Regge behavior for $x\rightarrow 0$ and duality ideas;
\item[iii)] invariance under charge conjugation and isospin.
\end{itemize}

These considerations define in the case of the valence quarks the following
structure function \cite{Altarelli1}

\bq
\Phi_{qq_v}({x}, \mu_0^2)
= { \Gamma(A + {1 \over 2}) \over 
\Gamma({1 \over 2}) \Gamma(A) }
{ (1-x)^{A-1} \over \sqrt{x} }.
\label{csf1}\eq
For the sea quarks the corresponding structure function becomes

\bq
\Phi_{qq_s}({x}, \mu_0^2)
= { C \over x } (1-x)^{D-1},\label{csf2}
\eq
and in the case of the gluons we take

\bq
\Phi_{qg}({x}, \mu_0^2)
= { G \over x } (1-x)^{B-1}~.\label{csf3}
\eq

The other ingredients of the formalism, i.e., 
the probability distributions for 
each constituent quark, are defined according to the procedure of Traini
et al. \cite{Traini}, that is, a constituent quark, $q_0$, has a probability
distribution determined by

 \bq
q_0(z,\mu_0^2) =  \frac{1}{(1-x)^2} \int d^3k n_q (|\vec{k}|)
\delta\left(\frac{z}{1-z}-\frac{k^+}{M}\right),
\label{parton1}
\eq
where $n_q (|\vec{k}|)$ is the
momentum  density distribution for the constituent quark momentum $\vec{k}$ 
in the corresponding baryonic state and can be calculated as

\bq
n_{u/d}= \;\;<N|\sum_{i=1}^3\delta{(\vec{k}-\vec{k}_i)} 
\frac{1 \pm \tau^z_i}{2}|N>.
\eq
Eq.(\ref{parton1}) includes support correction 
and satisfies the particle number sum rule \cite{Traini}.

Our last assumption relates to the hadronic scale $\mu_0^2$, i.e., 
that  at which the constituent quark 
structure is defined. We choose 
$\mu_0^2$, as defined in Ref. \cite{Traini}, namely by fixing the momentum 
carried by the various partons. This hypothesis determines {\it all 
 the parameters of
the approach (Eqs. (\ref{csf1}) through (\ref{csf3})), except one}. In fact,
the constants $A$, $B$, $G$
and the ratio $C/D$ are determined by the amount of momentum carried by the 
different partons, i.e. by the 2$^{nd}$ moments of the parton distributions. 
These quantities are experimentally known at high $Q^2$. 
We recover their values at the low $\mu_0^2$ scale by performing
a NLO backward evolution in the DIS scheme. 

The experience accumulated during the last years \cite{Traini} 
suggests the use of a hadronic scale, $\mu_0^2=0.34$ GeV$^2$. At this value of 
the momentum transfer perturbative $QCD$ at NLO tells us that
53.5 $\%$ of the nucleon momentum is carried by the valence quarks, 
35.7 $\%$ by the gluons and the rest by the sea.
This choice  of the hadronic scale fixes the parameters $A$, $B$, $G$ and 
the ratio $C/D$. Using besides some phenomenological input, the following 
parameters  have been obtained:
$A=0.435$, $B=0.378$, $C=0.05$, $D=2.778$ and $G=0.135$ \cite{Scopetta}.
We stress for later purposes that the unpolarized structure function 
$F_2$ is rather insensitive to the change 
of the sea ($C$, $D$) and the gluon ($B$, $G$) constants.

To complete the process \cite{Parisi-Petronzio,Glueck-Reya,Jaffe-Ross} the above input 
distributions are NLO-evolved in the DIS scheme to 10 GeV$^2$, where they 
are compared with the data.

We next generalize our previous discussion to the polarized case. The functions
$\Phi_{ab}$ now specify spin and flavor. We next construct the polarized parton
distributions with particular emphasis on spin. Let

\bq
\Delta q (x,\mu_0^2) = q_+ (x,\mu_0^2) - q_- (x,\mu_0^2)
\eq
where $\pm$ label the quark spin projections and $q$ represents any flavor. 
The generalized $ACMP$ approach implies 
\bq
q_i(x,\mu_0^2) = \int_x^1 \frac{dz}{z} \sum_j (u_{0j}(z,\mu_0^2) 
\Phi_{u_{j}q_i} (\frac{x}{z},\mu_0^2) + d_{0j}(z,\mu_0^2)
\Phi_{d_{j} q_j}(\frac{x}{z},\mu_0^2))
\eq
where $i=\pm$ labels the partonic spin projections and $j=\pm$ the constituent
quark spins. Using spin symmetry we arrive at \footnote{We omit writing
explicitly the hadronic scale dependence from now on, unless needed for
clarity.}

\bq
\Delta q (x) =\int_x^1 \frac{dz}{z} ( \Delta u_0 (z) \Delta
\Phi_{uq}(\frac{x}{z}) + \Delta d_0 (z)\Delta \Phi_{dq}(\frac{x}{z}))
\eq
where $\Delta q_0 = q_{0+} - q_{0-}$, and 

\bq
\Delta \Phi_{uq} = \Phi_{u+q+} - \Phi_{u+q-}
\eq
\bq
\Delta \Phi_{dq} = \Phi_{d+q+} - \Phi_{d+q-}
\eq
Note at this point that the unpolarized case can be described in this
generalized formalism as

\bq
 q (x) =q_+(x) + q_-(x) = 
\int_x^1 \frac{dz}{z} ( u_0 (z) \Phi_{uq} (\frac{x}{z})+ 
d_0 (z) \Phi_{dq}(\frac{x}{z})),
\eq
where

\bq
\Phi_{uq} = \Phi_{u+q+} + \Phi_{u+q-},
\eq
\bq
\Phi_{dq} = \Phi_{d+q+} + \Phi_{d+q-}.
\eq

We next reformulate the description in term of the conventional valence and sea
quark separation, i.e.,

$$\Delta q (x) =\Delta q_v (x) +\Delta q_s (x) = $$
\bq
\int_x^1 \frac{dz}{z} (\Delta u_0 (z) (\Delta \Phi_{u q_v}(\frac{x}{z}) 
+\Delta \Phi_{u q_s}(\frac{x}{z})) + \Delta d_0 (z) (\Delta \Phi_{d q_v} (\frac{x}{z})+ 
\Delta \Phi_{d q_s} (\frac{x}{z})) )
\eq
Here $\Delta \Phi_{{q} q_v} = \Delta \Phi_{q q_v} \delta_{{q} q}$,
therefore
\bq
\Delta q_v (x) =\int_x^1 \frac{dz}{z} \Delta q_0 (z) \Delta \Phi_{qq_v} 
(\frac{x}{z}),
\eq
\bq
\Delta q_s (x) = \int_x^1 \frac{dz}{z} (\Delta u_0 (z) \Delta \Phi_{uq_s} 
(\frac{x}{z}) + \Delta d_0 (z) \Delta \Phi_{d q_s} (\frac{x}{z})).
\label{sea}
\eq
We introduce $SU(6)$ (spin-isospin) symmetry as a simplifying assumption, which
leads to 

\bq
\Delta \Phi_{uu} = \Delta \Phi_{dd}
\eq
and

\bq
\Delta \Phi_{ud} = \Delta \Phi_{du}.
\eq
 Furthermore they imply 

\bq
\Delta \Phi_{u u_v} +  \Delta \Phi_{u u_s} = \Delta \Phi_{d d_v} +  \Delta
\Phi_{d d_s}
\eq
and 

\bq
\Delta \Phi_{u d_s} = \Delta \Phi_{d u_s}.
\eq
 
If we now add to these the following relations

\bq
\Delta \Phi_{u u_s} = \Delta \Phi_{d u_s},
\label{simp1}
\eq

\bq
\Delta \Phi_{d d_s} = \Delta \Phi_{u d_s},
\label{simp2}
\eq
which are beyond $SU(6)$ symmetry, but quite reasonable, we obtain the following
equalities

\bq
\Delta \Phi_{u u_s} = \Delta \Phi_{d u_s} = \Delta \Phi_{u d_s} = 
\Delta \Phi_{d d_s} = \Delta \Phi_{q q_s}
\label{phisea}
\eq
and

\bq 
\Delta \Phi_{u u_v} =\Delta \Phi_{d d_v} =\Delta \Phi_{q q_v}.
\eq

In this way we reduce the structure functions for the valence and for the sea to
just two independent constituent structure functions and Eq. (\ref{sea}) 
simplifies to

\bq
\Delta q_s (x) = \int_x^1 \frac{dz}{z} (\Delta u_0 (z)  + \Delta d_0 (z)) 
\Delta \Phi_{qq_s}(\frac{x}{z}).
\label{deltasea}
\eq

The same argumentation applied to gluons implies

\bq
\Delta g (x) = \int_x^1 \frac{dz}{z} (\Delta u_0 (z) + \Delta d_0 (z)) 
\Delta \Phi_{q g}(\frac{x}{z})
\eq
and we recover our old expression

\bq
g (x) = \int_x^1 \frac{dz}{z} (u_0 (z) + d_0 (z) )\Phi_{q g}(\frac{x}{z}).
\eq

We may conclude our analysis up to now by stating that the $ACMP$ procedure can
be extended to the polarized case just by introducing three additional
structure functions for the constituent quarks: $\Delta \Phi_{q q_v}$, $\Delta
\Phi_{q q_s}$ and $\Delta \Phi_{q g}$.

In order to determine the polarized constituent structure functions we add some
assumptions which will tie up the constituent structure functions for the
polarized and unpolarized cases completely, reducing dramatically the number of
parameters. They are:

\begin{itemize}

\item [iv)] factorization assumption: $\Delta \Phi$ cannot depend upon the quark 
model used, i.e, cannot depend upon the particular $\Delta q_0$;

\item [v)] positivity assumption: the positivity constraint $\Delta \Phi \leq
\Phi $ is saturated for $x = 1$.

\end{itemize}

We next discuss how these additional assumptions determine completely the
parameters of the polarized constituent structure functions and discuss the
physics implied by them.

The $QCD$ partonic picture, Regge behavior and duality imply that

\bq
\Delta\Phi_{q f} = \frac{\Delta C_f (1-x)^{A_f - 1}}{x^{a_f}}
\eq
and $ -\frac{1}{2}< a_f < 0$, for all $f= q_v, q_s, g$, as defined by dominant 
exchange of the $A_1$ meson trajectory \cite{Heimann-Ellis}.

The positivity restriction, $\Delta \Phi \leq \Phi$, is a natural consequence of
the probability interpretation of the parton distributions. The assumption that
the inequality is saturated for $x=1$, in the spirit of ref. \cite{kaur}, 
implies that $\Delta C_f = C_f$, the latter being the parameter fixed in the
analysis of the unpolarized case, and therefore when $x \approx 1$ the
partons which carry all of the momentum also carry all of the polarization,
i.e., $\Phi_{+-} = 0$. From the point of view of the number of parameters, this
is a minimalistic assumption, since it reduces the parameters of the polarized
case to those of the unpolarized one. Lastly the exponents $A_f$ are also taken
from the unpolarized case. To summarize the parametrization, let us stress 
that the change between the polarized
functions and the unpolarized ones comes only from Regge behavior.

Let us insert here a comment about the constituent quark structure functions.
Under the conditions imposed by
the generalized $ACMP$, namely that 
the low $x$ regime is governed by the Regge behavior,
and the 
large $x$ behavior 
is determined by assuming soft
partons to be independently emitted 
\cite{Altarelli1}, the $\Phi$ functions are
of the form

\bq
\Phi_{q+ q\pm} (x) = \sum_i \frac{C_{i\pm}}{x^{a_i}}(1-x)^{A-1},
\eq
where $i$ sums over leading trajectories. That Regge behavior 
works reasonably well {\sl at scales typical of
soft hadronic physics} has been recently confirmed \cite{abfr}. 
For valence quarks, for example,
the sum is limited to the rho meson ($a_1 = \frac{1}{2}$) and the $A_1$-meson 
($ -\frac{1}{2}\leq a_2 \leq 0$). 
The unpolarized structure function is given  then by

\bq
\Phi_{qq_v} = \left(\frac{C_{1+} + C_{1-}}{x^{a_1}} + \frac{C_{2+} +
C_{2-}}{x^{a_2}}\right)(1-x)^{(A-1)}
\eq
and  the polarized one by 

\bq
\Delta \Phi_{qq_v} = \left(\frac{C_{1+} - C_{1-}}{x^{a_1}} + \frac{C_{2+} -
C_{2-}}{x^{a_2}}\right)(1-x)^{(A-1)}.
\eq
The observed Regge behavior implies
\bq
C_{1+} = C_{1-}\;\;\;,\;\;\; C_{2+} = - C_{2-}\,,
\eq
and therefore the shapes used above arise. Moreover our additional
assumption for the large $x$ limit \cite{kaur} leads to

\bq
C_{1+} = C_{2+}\,.
\eq
Similar arguments hold for the sea and the gluons.

With all the above hypothesis our constituent quark functions become

\bq
\Delta \Phi_{qq_v}({x}, \mu_0^2)
= { \Gamma(A + {1 \over 2}) \over 
\Gamma({1 \over 2}) \Gamma(A) } x^{\alpha}
{ (1-x)^{A-1} } 
\label{pi}
\eq
\bq
\Delta \Phi_{qq_s}({x}, \mu_0^2)
= C  x^{\alpha}
{ (1-x)^{D-1} } 
\eq
\bq
\Delta \Phi_{qg}({x}, \mu_0^2)
= G  x^{\alpha}
{ (1-x)^{B-1} }
\label{pf} 
\eq
where $A,C,D,B,G$ are the parameters of the unpolarized case. In what follows,
we shall use
$0 \leq \alpha \leq0.5$, the range proposed by ref. \cite{abfr} 
in agreement with \cite{Heimann-Ellis}.

The other ingredients of the formalism, i.e., 
the probability distributions for 
each constituent quark, are defined according to the procedure of Traini
et al. \cite{Traini}, that is, a constituent quark, $q_{0\pm}$, has a 
probability distribution determined by

 \bq
q_{0\pm}(z,\mu_0^2) =  \frac{1}{(1-x)^2} \int d^3k n_{q_{0 \pm}} (|\vec{k}|)
\delta\left(\frac{z}{1-z}-\frac{k^+}{M}\right),
\label{partons1}
\eq
where $q_{0\pm}$ denotes the $q_0$th constituent quark whose spin is aligned
(anti-aligned) to the nucleon´s spin while $n_{q_{0\pm}} (|\vec{k}|)$ is the
momentum  density distribution for the valence quark momentum $\vec{k}$ 
and equivalent spin projection. $n_{q\pm} (|\vec{k}|)$ can be evaluated
projecting out the appropriate spin and flavor component of the constituent
quark and in the corresponding baryonic state is given by \cite{Traini}

\bq
n_{u_0/d_0 , \pm}= \;\;<N ,\; J_z = +\frac{1}{2}|
\sum_{i=1}^3\delta{(\vec{k}-\vec{k}_i)} 
\frac{1 \pm \tau^z_i}{2}\frac{1 \pm\sigma^z_i}{2}|N ,\; J_z = +\frac{1}{2}>
\eq
Eq.(\ref{partons1}) includes support correction and satisfies the particle
number
sum rule \cite{Traini}.

Let us briefly comment about the other leading twist polarized structure
function, namely the chiral odd transversity function $h_1$.
The question we briefly want to address is how the $ACMP$ procedure changes our
previous conclusions \cite{Scopetta1}. Kirschner et al. \cite{Kirschner} find 
that the Regge behavior for $h_1$ is roughly constant and therefore consistent 
with our choice for the behavior of $g_1$, i.e., $\sim x^{0 \div \frac{1}{2}}$
\cite{Heimann-Ellis}. Note that the Regge behavior is dominant at low $x$
and low $Q^2$, where the enhancement due to gluon radiation, given by QCD
evolution, is not yet efficient.

The determination of the large $x$ behavior of the structure function as
discussed previously for $g_1$ is dominated by the independence of the softly
emitted partons and therefore should be the same for the transversity
distributions. These arguments lead us to conjecture that, for
non-relativistic models of hadron structure, the $ACMP$ mechanism mantains at
the hadronic scale the almost equality between these two polarized structure
functions,

\bq
h_1(x,\mu_0^2) = g_1(x,\mu_0^2).
\eq
Evolution will produce the difference between them as already stated in our
previous analysis \cite{Scopetta1}.
 
\section{Results}
We will discuss the results of our calculation 
of the polarized structure function $g_1$ for the
proton and the neutron,
evaluating the polarized constituent momentum distributions,
$\Delta u_o$ and $\Delta d_o$, within
the algebraic model of Bijker, Iachello and Leviatan
\cite{Bijker}, which proved so succesful in our previous work \cite{Scopetta}.
The parameters of the wave functions are kept as determined by
their authors, which fitted them to static properties of hadrons, since
the present scheme provides us automatically with the momentum sum rule
and therefore no ad hoc modification of the model wave functions has been
necessary.

To evaluate $g_1$ at the experimental scale we will perform a NLO evolution of
the model parton distributions. It is known that perturbative $QCD$ to this
order allows the proposal of varied factorization schemes \cite{fp},
whereby one is able to define the partonic distributions in 
different ways without altering the physical observables. In our analysis the
parton distributions are determined by the quark model through the ACMP
procedure. Which factorization scheme should we apply? It is evident that
different factorization schemes lead to different results. We have adopted on
physical grounds the $AB$ scheme as defined in ref.\cite{bfr} \footnote{It
consists in modifying minimally the NLO polarized anomalous dimensions 
\cite{NLOP}, calculated in the $\overline{MS}$ scheme, 
in order to have the axial anomaly governing the first moment of $g_1$, 
as proposed in Ref \cite{Altarelli-Ross}.}. In it some relevant physical 
observables, such as $\Delta \Sigma$,  are scale independent, i.e., they behave as conserved quantities, 
and therefore the partons have a well defined interpretation in terms of 
constituents \footnote{It must be pointed out that, if we had used  
the  $\overline{MS}$ scheme, the results would have been in better agreement 
with the data.
However we believe that this accidental agreement hides some physics, as 
will become clear later on.}.
 A detailed analysis of factorization scheme dependence in model
calculations of polarized structure functions will be presented elsewhere
\cite{tbp}.

Let us initially use our previous parametrization of the $\Delta \Phi$ 
functions as
determined in the unpolarized case, Eqs.  (\ref{pi}) -- (\ref{pf}),
with all the caveats expressed repeatedly
\cite{Scopetta}. 
Then we are able to predict the parton distributions at the hadronic 
scale and therefore their first moments can be calculated, leading to:

$$\Delta  q_v (\mu_0^2)  =  \Delta  u_v (\mu_0^2)
+ \Delta  d_v (\mu_0^2) = $$ 
$$ = \int dz ( \Delta u_0(z,\mu_0^2)
+ \Delta d_0(z,\mu_0^2) ) \int dx \Delta \Phi_{q q_v}(x)
$$
\bq
 =   0.534 \div 0.662~,
\eq
the first value corresponding to $\alpha =0.5$ and the second to $\alpha = 0$,
being $\alpha$ the Regge intercept;
\bq
\Delta  q_s (\mu_0^2) = \int dz ( \Delta u_0(z,\mu_0^2)
+ \Delta d_0(z,\mu_0^2) ) \int dx \Delta \Phi_{q q_s}(x) = 0.0085 \div 0.018,
\label{rsea}
\eq
which disagrees not only in magnitude, but also in sign,
with the AB data analysis in \cite{abfr}. In fact one should realize
that
our calculation contains certain peculiarities 
since we 
use the AB evolution scheme in a symmetric sea, a feature which is in the 
spirit of the ACMP model.
The structure of our sea, as reflected in Eqs. (\ref{phisea}) and
(\ref{deltasea}) leads to simplifications in the evolution. In particular, 
the sea does not contribute to 
$a_3$ nor $a_8$ ,i.e., under our assumptions the different contributions 
cancel out. Keeping this in mind, it turns out that $\Delta q_s(\mu_0^2)$ is
related to $\Delta s$, the first moment of the polarized strange sea, and using
for the latter the number obtained in \cite{abfr} one gets: $\Delta  q_s
(\mu_0^2) 
= {1 \over 2} \Delta s
= -0.022 \pm 0.005$, at variance with (\ref{rsea}).

We can also predict:

\bq
\Delta  g (\mu_0^2)= \int dz ( \Delta u_0(z,\mu_0^2)
+ \Delta d_0(z,\mu_0^2) ) \int dx \Delta \Phi_{q g}(x) = 0.295 \div 0.357,
\label{g}
\eq
in reasonable agreement with the recent calculation of
ref. \cite{barone}.

Using our assumptions for the sea, Eqs. (\ref{phisea}) and (\ref{deltasea}), 
we may calculate the following quantities:
\bq
a_8 = \Delta  u_v (\mu_0^2) + \Delta  d_v (\mu_0^2) =
0.534 \div 0.662~,
\label{a8}
\eq
i.e., the octet charge, whose experimental value is $0.58\pm0.03$ 
\cite{cr},
and
\bq
\Delta  \Sigma = 6 \Delta  q_s (\mu_0^2) + a_8 =
0.584 \div 0.770 \, , \eq
which is our prediction for the spin carried by the quarks and the antiquarks.
The estimate for this quantity in \cite{abfr} is $0.45 \pm 0.09$. 

For the singlet charge we get
\\
\bq
 a_0 (\mu_0^2) = 
\Delta  \Sigma - 3 {\alpha_s(\mu_0^2) \over 2 \pi} \Delta g (\mu_0^2) =
0.511 \div 0.682 \, , \eq
and finally, for the isospin charge  
\bq
a_3 =\Delta u - \Delta d = 0.888 \div 1.104~,
\label{a3}
\eq
to be compared with $a_3 = 1.257 \pm 0.003$, experimental
value as in \cite{pdg}.

In order to compare with the experimental DIS data we perform now the AB 
evolution.
Results are shown in Figs. 1, 2 for the proton and the neutron, respectively.
It sems that some physics 
is missing in the above description as signalled by the disagreement with 
the neutron data. 

Looking at the first moments we realize that specially 
our determination for the sea looks very poor (cf. Eq.(\ref{rsea}) and
discussion below.). 
Can this be the origin of the neutron problem?
The fact that our result (\ref{rsea}) is inadequate 
means that our hypothesis (v), 
based on ref. \cite{kaur}, does not apply to the sea.

Let us choose $\Delta \Phi_{q q_s}$ to reproduce the experimental
sea data. This implies 

\bq
\Delta \Phi_{q q_s}({x}, \mu_0^2)
= \Delta C_s x^{\alpha}
{ (1-x)^{D-1} } 
\label{ns}
\eq

where $\Delta C_s = -0.13 \div -0.06$, which has been chosen so that

\bq
 \Delta  q_s (\mu_0^2) =  -0.022 \pm 0.005
\eq
in agreement with ref. \cite{abfr}. 

With this parametrization we also get
\bq
a_0 (\mu_0^2) = 0.330 \div 0.443~,
\eq
and our prediction for the spin carried by the constituents is 
\bq
\Delta  \Sigma = 0.402 \div 0.530~,
\eq
being $0.45\pm0.09$ the value given in \cite{abfr}.  
Finally, the estimates for $\Delta g(\mu_0^2)$, $a_8$, and $a_3$ do not change 
and are given by Eqs. (\ref{g}), (\ref{a8}) and (\ref{a3}).
\\

After evolving, Fig.3 shows that the neutron calculation improves 
substantially, 
while the proton one remains quite the same as shown in Fig. 4. It is thus 
clear that it was too naive to use the unpolarized data to fit the 
polarized constituent sea structure function. 

We have traced back the remaining disagreement with the neutron data to the
symmetric treatment of $u$ and $d$ quarks.
It can be shown that a weak breaking of the SU(6) symmetry in the quark 
model  used \cite{Bijker} and/or in the constituent quark structure functions 
improves considerably the agreement \cite{tbp}. The neutron is extremely
sensitive to small  changes in the
valence structure.

\section{Conclusions and Outlook}

The present calculation shows that our description of
the unpolarized structure functions \cite{Scopetta} can be succesfully
extended to the polarized case. Therefore,
low energy models seem to be  consistent with DIS data when a structure
for the constituent quarks is introduced.
We have chosen this structure, 
following the ACMP description \cite{Altarelli1}, 
to be consistent with well known phenomenological
inputs, such as Regge behavior at low $x$ and counting rules at large $x$.
This assumption made it possible in the unpolarized case to fully define the
procedure with only one new parameter, to which the predictions where not very
sensitive.

Using a physically motivated minimal prescription for the polarized case, 
with no additional parameters, we are able to obtain a good prediction
of the the proton data \cite{emc,smc}. The minimal procedure fails, however, 
to reproduce  the recent accurate neutron data \cite{e154}.
Relaxing the minimal procedure, with the addition of only {\bf one} new
parameter to define the polarized sea, we obtain a 
significantly improved description also for the neutron data.
A perfect agreement would be gotten by introducing an SU(6) breaking 
mechanism. It is worth emphasizing the importance of the neutron data in
disentangling the fine details of the structure.

The outcome of our calculation is not surprising. 
We had defined the sea by looking at observables, like the
unpolarized structure function $F_2$
\cite{Scopetta}, which are not very sensitive to scrutinize its
distributions. Here, we are analyzing new polarization observables, 
which depend stronger on the structure of the sea. This new information
has motivated a more precise determination of the constituent quark structure
functions. For example, it is well known, that the
Ellis-Jaffe sum rule for the neutron arises from the vanishing of the polarized
sea. This description had to be abandoned when the EMC data first appeared. 
Showing that a not negligible $g_1$ for
the neutron is consistent with a large contribution from the polarized sea is
just rediscovering history in our scheme. 

The calculation has also clarified the role of the gluons and the valence
quarks. It is clear that the gluons become important through the evolution
process, i.e., it is the soft bremsstrahlung gluons which acquire a large
portion of the partonic spin. Moreover in the case of the neutron we have
realized that the SU(6) symmetric valence quark contribution basically cancels
out and it is the deviation from this symmetry which leads to our remaining
discrepancy.

The crucial role played by the sea and isospin breaking effects in
the description of the polarized data is demanding an explanation beyond our
solution. As mentioned in other occasions, our starting quark model does not
implement chiral symmetry breaking, 
therefore the sea is generated solely at the
level of our constituent quark structure functions. Does the procedure thus far
developed implement chiral symmetry breaking properly? Data seem to confirm 
this statement. However, there is an alternative approach, originating also at
the very beginning of DIS parton physics \cite{kuti}, which has gained 
followers after the rebirth of effective theories and we could
label under the generic name of chiral procedure.
Under this philosophy, the sea would originate also from the mesonic degrees 
of freedom used to define the chiral quark models. One expects, that the 
factorization procedure developed in our approach to incorporate 
the constituent quark structure, could be extended to these models by 
introducing also the structure functions of the elementary mesons. It has to be
investigated if this scheme is able to produce as good descriptions of the data
as the present one. If this were the case, there would be
a duality of approaches modelling the confinement mechanism.
To test these models, until $QCD$ is not solved, experiments should guide 
the efforts with the aim of predicting  new phenomenology.

A comment about the most common factorization schemes is necessary. Had we
chosen the $\overline{MS}$ scheme for our calculation, our result would have
basically fitted the data. The sea would have been determinant in explaining the
data and the other mentioned effects practically inexistent. Taking the AB
scheme has led to good, but certainly not perfect, agreement with the data and
therefore to the necessity of finding other possible mechanisms to
interpret the nucleonic structure. It is important that in sensitive scenarios,
like the neutron, one uses a completely consistent procedure.

We would like to stress that within
our procedure the {\it spin problem}, as initially presented,  does not arise. 
The constituent quarks carry all of the
polarization. When their structure is unveiled this polarization is split among
their different partonic contributions 
in the manner we have described and which
is consistent with the data.
 The quality of both unpolarized and polarized data thus 
far analyzed confirm the
validity of the approach. We have showed also, that with very reasonable
assumptions, the scheme becomes highly predictive, a feature which is necessary
for the planning of future experiments.

\section*{Acknowledgements}

We have benefitted greatly from the constructive criticism of the referees. In
particular, one of them pointed out that we had taken some input parameter in 
the wrong factorization scheme, and therefore our previous results 
overemphasized the role of the sea in the explanation of the data.

\newpage
\centerline{\bf \Large Captions}
\hfill\break
\hfill\break
\hfill\break
{\bf Figure 1}: We show the structure function $xg_1^p(x,Q^2)$ obtained 
at $Q^2 = 10$ GeV$^2$ by evolving at NLO the model calculation \cite{Bijker} 
without considering the structure of the constituents \cite{Traini} (dashed). 
The same quantity, $xg_1^p (x,Q^2)$, evolved at NLO to 
$Q^2=10$ GeV$^2$, for the two extreme Regge behaviors
mentioned in the text, is given by the two full curves 
($\alpha =0$ is the upper curve, here and also in the following figures). 
The parameters used are those of our calculation for the unpolarized case
\cite{Scopetta}.
The data from refs. \cite{emc,smc} at $Q^2 \approx 10$ GeV$^2$ are also shown.

\hfill\break
\hfill\break
{\bf Figure 2}: The structure function  $xg_1^n(x,Q^2)$ for the neutron
evolved at NLO 
to
$Q^2= 5$ GeV$^2$, for the two extreme Regge behaviors
mentioned in the text, is shown by the two full curves. 
The parameters used are those of our calculation for the unpolarized case
\cite{Scopetta}.
The data from refs.
\cite{e154} at $Q^2= 5$ GeV$^2$ are also shown.

\hfill\break
\hfill\break
{\bf Figure 3}: The structure function $xg_1^n(x,Q^2)$ for the neutron
evolved at NLO 
to
$Q^2=5$ GeV$^2$, for the two extreme Regge behaviors
mentioned in the text, is shown by the two full curves. 
The new parametrization of the sea, Eq. (\ref{ns}), is used.
The data from refs.
\cite{e154} at $Q^2 = 5$ GeV$^2$ are also shown.

\hfill\break
\hfill\break
{\bf Figure 4}: The structure function $xg_1^p (x,Q^2)$ for the proton
evolved at NLO to
$Q^2=10$ GeV$^2$, for the two extreme Regge behaviors
mentioned in the text, is given by the two full curves. 
The new parametrization of the sea, Eq. (\ref{ns}), is used.
The data from refs. \cite{emc,smc} at $Q^2 \approx 10$ GeV$^2$ are also shown.

\newpage
\begin{figure}[h]
\vspace{12cm}
\includegraphics{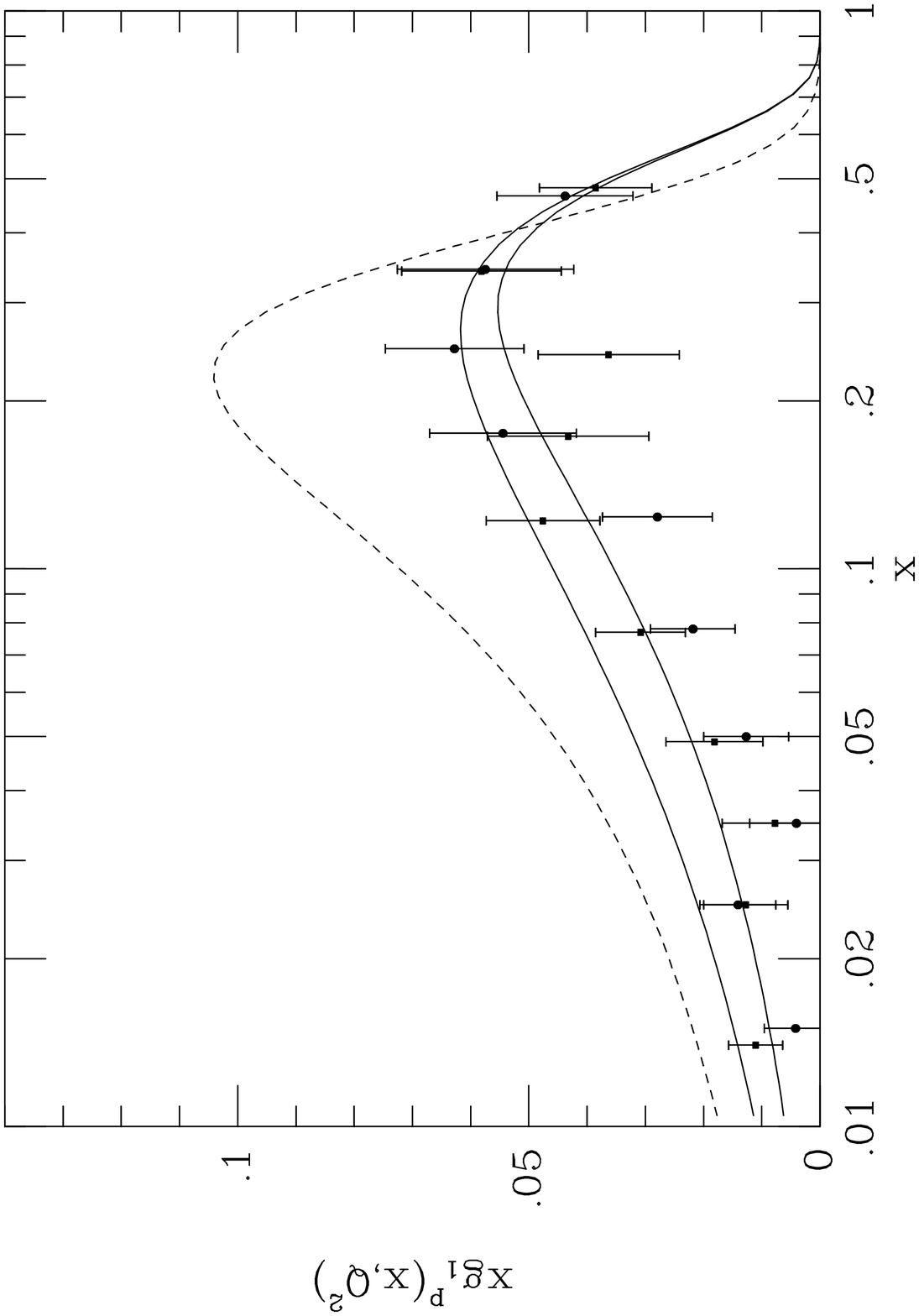}
\end{figure}
\vspace{3cm}
\centerline{\large S. Scopetta, V. Vento and M. Traini 
}
\vspace{1cm}
\centerline{\bf \large FIGURE 1}

\newpage
\begin{figure}[h]
\vspace{12cm}
\includegraphics{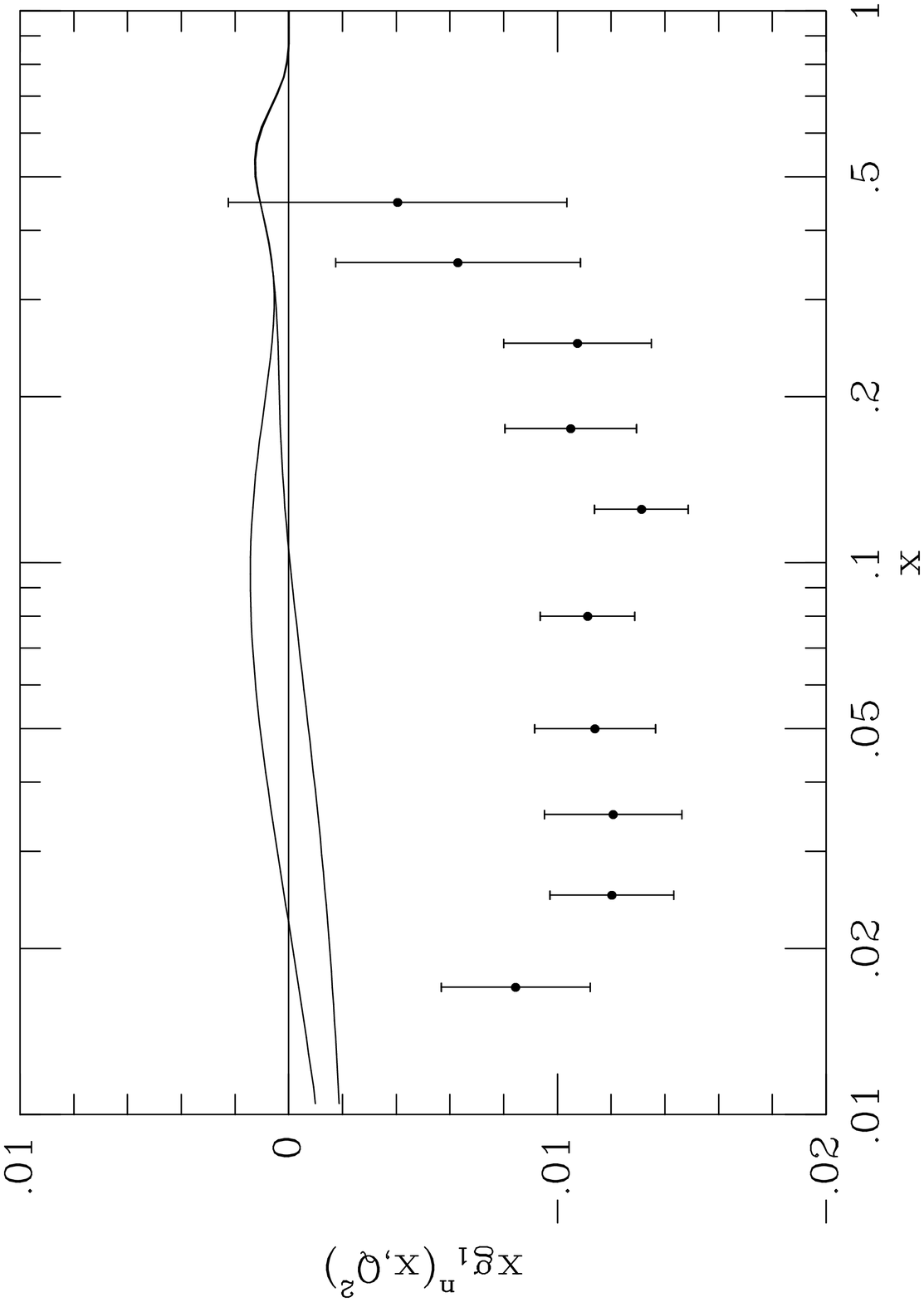}
\end{figure}
\vspace{3cm}
\centerline{\large S. Scopetta, V. Vento and M. Traini
}
\vspace{1cm}
\centerline{\bf \large FIGURE 2}

\newpage
\begin{figure}[h]
\vspace{12cm}
\includegraphics{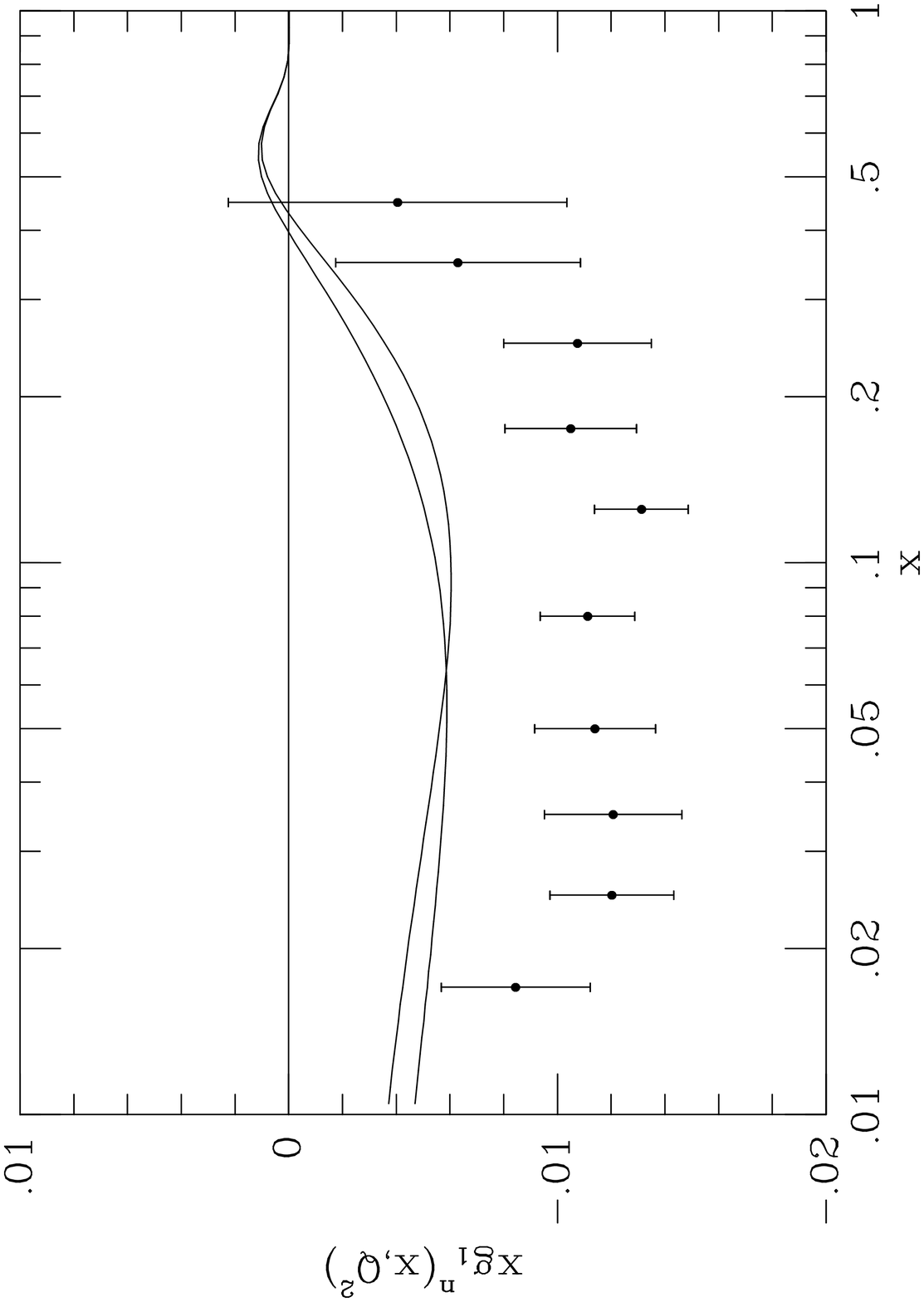}
\end{figure}
\vspace{3cm}
\centerline{\large S. Scopetta, V. Vento and M. Traini 
}
\vspace{1cm}
\centerline{\bf \large FIGURE 3}

\newpage
\begin{figure}[h]
\vspace{12cm}
\includegraphics{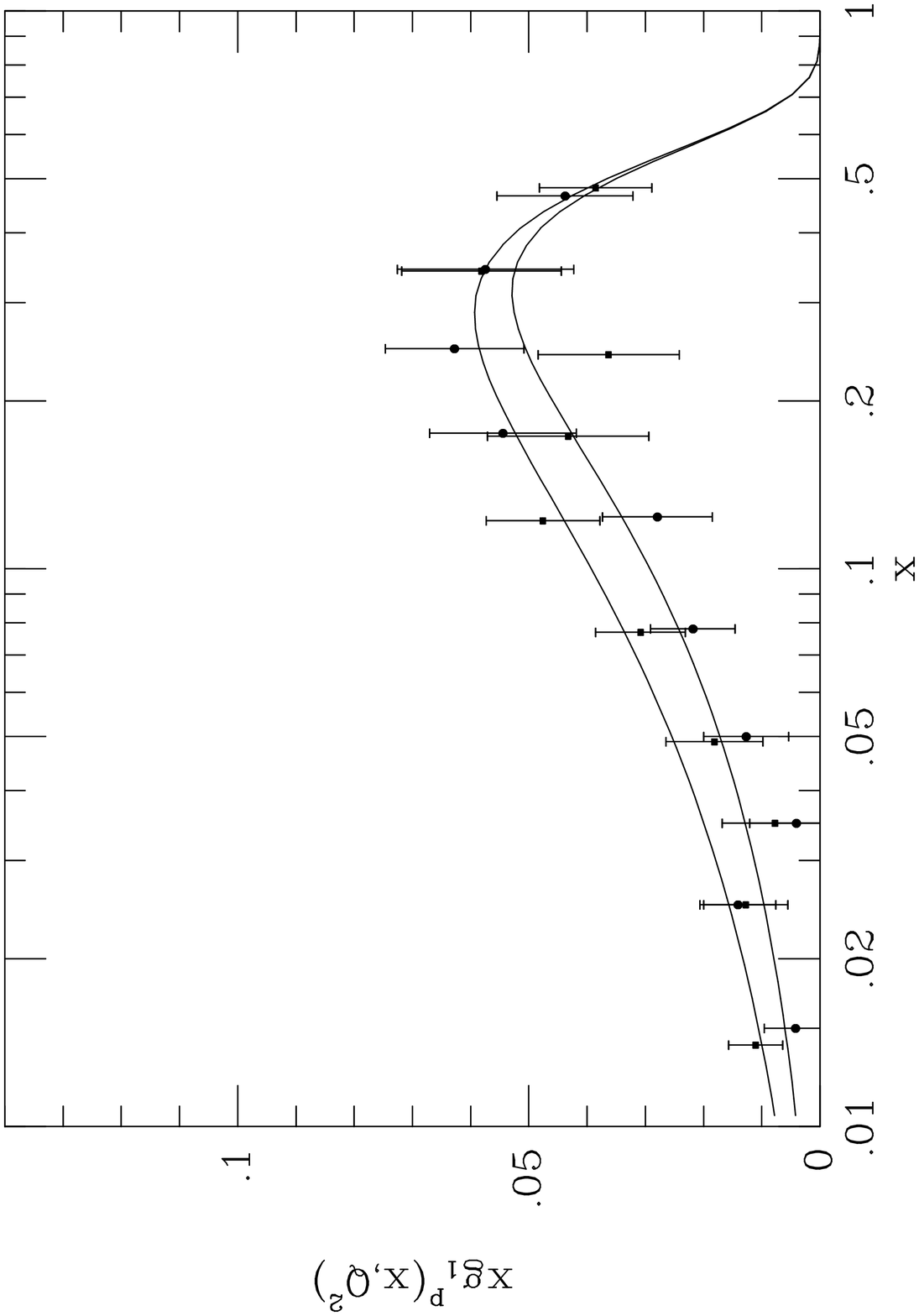}
\end{figure}
\vspace{3cm}
\centerline{\large S. Scopetta, V. Vento and M. Traini
}
\vspace{1cm}
\centerline{\bf \large FIGURE 4}

\end{document}